\def\nat{{\tt I\kern-.2em{N}}}
\def\hyper#1{\ ^*\kern-.2em{#1}}
\def\hypernat{{^*{\nat }}}
\def\eskip{\hskip.25em\relax} 
\def\Hyper#1{\hyper {\eskip #1}}
\def\power#1{{{\cal P}(#1)}}
\def\iff{\leftrightarrow}
\def\qed{{\vrule height6pt width3pt depth2pt}\par\medskip}
\def\m@th{\mathsurround=0pt}

\def\id{\par\hangindent2\parindent\textindent}
\def\textindent#1{\indent\llap{#1}}
\magnification=\magstep1
\tolerance 10000
\baselineskip  12pt
\hoffset=.25in
\hsize 6.00 true in
\vsize 8.85 true in
\font\eightrm=cmr9
\centerline{\bf Standard and Hyperfinite Unifications for Physical Theories}\par\bigskip 
\centerline{Robert A. Herrmann}\par\medskip
\centerline{Mathematics Department}
\centerline{U. S. Naval Academy}
\centerline{572C Holloway Rd.}
\centerline{Annapolis,  MD 21402-5002}
\centerline{15 FEB 2001, last revision 25 APR 2010}\bigskip
{\leftskip=0.5in \rightskip=0.5in \noindent {\eightrm {\it Abstract:} A set of physical theories is represented by a nonempty subset $\rm \{S^V_{N_j}\mid j \in \nat \}$ of the lattice of consequence operators defined on a language ${\Lambda}.$ It is established that there exists a unifying injection $\cal S$ defined on the nonempty set of significant representations for natural-systems $\rm M \subset \Lambda.$ If $\rm W \in M,$ then ${\cal S}_{\rm W}$ is a hyperfinite ultralogic and $\bigcup\{{\bf S}^{\bf V}_{{\bf N}_j}({\bf W})\mid j \in \nat \} = {\cal S}_{\rm W}(\Hyper {\bf W})\cap {\bf \Lambda}.$ A ``product'' hyperfinite ultralogic $\Pi$ is defined on internal subsets of the product set $\hyper {\bf \Lambda}^m$ and shown to represent the application of $\cal S$ to $\rm \{W_1,\ldots,W_m\} \subset M.$   There also exists a standard unifying injection $\rm S_W$ such that ${\cal S}_{\rm W}(\Hyper {\bf W})\subset \Hyper {\bf S_W}(\Hyper {\bf W}).$
 \par}}\par\bigskip
\noindent{\bf 1. Introduction.}\par\medskip
For reader convince, some of the introductory remarks that appear in Herrmann (2001a) are repeated. Seventy years ago, Tarski (1956, pp. 60-109) introduced {\it consequence} operators as models for various aspects of human thought. Within lattice theory, there are two such mathematical theories investigated, the {\it general} and the {\it finitary} consequence operators (Herrmann, 1987). The finitary consequence operators are usually the operators that model human thought processes that use but finite arguments and a finite collection of premises to arrive at a specific conclusion. Let $\rm L$ be a nonempty language, $\cal P$ be the power set operator and $\cal F$ the finite power set operator. \par\medskip
{\bf Definition 1.1.} A mapping $\rm C\colon \power{\rm L} \to \power {\rm L}$ is a general consequence operator (or closure operator) if for each $\rm X,\ \rm Y \in 
\power {\rm L}$\par\smallskip
\indent\indent (1) $\rm X \subset C(X) = C(C(X)) \subset L$; and if\par\smallskip
\indent\indent (2) $\rm X \subset Y$, then $\rm C(X) \subset C(Y).$\par\smallskip
\noindent A consequence operator C defined on L is said to be {\it finitary} ({\it finite}, or {\it algebraic}) if it satisfies\par\smallskip
\indent\indent (3) $\rm C(X) = \bigcup\{C(A)\mid A \in {\cal F}(\rm X)\}.$\par\medskip
{\bf Remark 1.2.} The above axioms (1), (2), (3) are not independent. Indeed, 
(1), (3) imply (2). Hence, the finitary consequence operators defined on a specific language form a subset of the general operators. The phrase ``defined on L'' means formally defined on $\power {\rm L}.$ \par\medskip
Natural-systems are named and defined by scientific disciplines. Each is an arrangement of named physical objects that are so related or connected as to form an identifiable unity. Except for the most basic, natural-systems always require the existence of accepted natural laws or processes for, at least, two events to occur. It is required that a natural-system either be constructed by application of natural laws or processes from more fundamental physical objects (natural-systems); or that the natural-system is altered in its development by such natural laws or processes, in which case the original natural-system may be considered as a more fundamental physical object. \par\smallskip 
Explicit statements for a natural law or process and the theories they yield are human inventions that imitate, in the words of Ferris (1979, p. 152), intrinsic natural laws or processes that govern the workings of those portions of our universe that are comprehensible. Individuals apply various mental processes to a set of hypotheses that include a set of natural laws or processes and predict behavior for a natural-system. Mental processes are also applied to natural laws or processes in order to construct our material ``man made universe.'' Consequence operators model such mental behavior. Indeed, these operators model many general mental processes not merely the standard notion termed as ``deduction.'' \par\medskip
\noindent {\bf 2. Axiomatic consequence operators.}\par\medskip
Prior to simplification, it is assumed that our consequence operators are axiomatic, where the axioms include appropriate natural laws or processes. Also, utilized is the fundamental philosophy of modern science notion that,  with the exception of the accepted and most fundamental of physical objects,  all named natural-systems are obtained by application of natural laws or processes to physical objects that are defined as more fundamental in character than the natural-systems of which they are constituents. Obviously, specified natural laws or processes alter specific natural-system behavior. As mentioned, the results in this paper are not restricted to what is usually termed as deduction. As done in Herrmann (1999, p. 12), only consider equivalent representatives as the members of $\rm L$. (This is not the same notion as consequence operator logical equivalence.) Let $\rm {\cal C}(L)$ [resp. $\rm {\cal C}_f(L)$] be the set of all general [resp. finitary] consequence operators defined on $\rm L,$ where $\rm A \subset \rm L$ is the set of logical axioms for $\rm F \in \rm {\cal C}(L)$ [resp. $\rm {\cal C}_f(L)$].\par\smallskip 

Although, usually,  such consequence operators are considered as axiomatic,  in this application the use of axiomless operators (Herrmann  1987, p. 3) leads to a significant simplification. For $\rm F \in \rm {\cal C}(L)$ [resp. $\rm {\cal C}_f(L)$], let $\rm A \cup \rm N\subset \rm L$ and suppose that  $\rm F(\emptyset)\supset \rm A \cup \rm N.$ (Note: $\rm N$ does not denote the natural numbers). Then  $\emptyset \subset \rm A \cup \rm N$ yields $\rm F(\emptyset) \subset \rm F(\rm A \cup \rm N),$ and $\rm A \cup \rm N \subset \rm F(\emptyset)$ yields that   $\rm F(\rm A \cup \rm N) \subset \rm F(\rm F(\emptyset)) = \rm F(\emptyset)$. Hence,  $\rm F(\emptyset) = \rm F(\rm A \cup \rm N).$ Further,  note that if $\rm B \subset \rm A \cup \rm N,$ then since $\emptyset \subset \rm B,$ it follows that $\rm F(\emptyset) = \rm F(\rm A \cup \rm N)\subset \rm F(\rm B)\subset \rm F(\rm F(\rm A \cup \rm N)) = \rm F( \rm A\cup \rm N)$ and $\rm F(\rm B) = \rm F(\rm A\cup \rm N).$  The objects in $\rm F(\rm A \cup \rm N)$ behave as if they are axioms for $\rm F.$ Can this axiomatic behavior be used to generate formally a specific consequence operator $\rm C,$ where $\rm C(\emptyset) = \emptyset,$ and the only results displayed by this model are conclusions not members of $\rm F(\rm A \cup \rm N)$? If such a meaningful consequence operator exists,  then this approach is acceptable since if natural laws or processes, as represented by $\rm N,$ are stated correctly,  such as always including any physical circumstances that might restrict their application,  then they behave like physical ``tautologies'' for our universe. For such a basic consequence operator $\rm F,$ the set $\rm F(\emptyset)$ is composed of all of the restatements of $\rm N$ that are considered as ``logically'' equivalent,  and all of the pure ``logical'' theorems.\par\smallskip 

 In general,  various forms of scientific argument are modeled by consequence operators,  where the use of axioms is a general process not dependent upon the axioms used. The axioms are but inserted into an argument after which the actual rules of inference are applied that might yield some $\rm x \in {\rm L} - \rm F(\emptyset)$. It is this $\rm x$ that may yield something not trivial. In the physical case,  this $\rm x$ may represent some aspect of an actual physical object distinct from the natural laws or processes.\par\medskip
\noindent {\bf 3. Extended Rules that generate consequence operators.}\par\medskip
In Herrmann (2001b), the unification presented is rather restrictive and does not properly represent common practice. A major aspect of a theoretical scientific construct is not usually included within the specific rules of inference for consequence operator generation, but it is an exterior notion. This does not mean that this aspect cannot be modeled by a consequence operator. However, in general, the composition of consequence operators does not lead to a consequence operator (Herrmann, 1987). This aspect is an extension of the realism relation. It is not adjoined to the axioms of a consequence operator but it is a basic notion within human intelligence and is applied only after a consequence operator generates conclusions from a set of premises. The definition that appears in the next section requires an extension of the usual rules of inference, an extension that is used in practice.\par\smallskip 
The following two paragraphs define ``logic-systems.'' In this investigation,  the term ``deduction'' is broadly defined. Informally,  the pre-axioms $\rm A\cup \rm N$ is a subset of our language ${\rm L},$ where $\rm N$ represents natural laws or processes, and there exists a fixed finite set ${\bf RI} =\{\rm  R_1,\ldots, R_p\}$ of n-ary relations $(\rm n \geq 1)$ on ${\rm L}.$ The term ``fixed'' means that no member of $\bf RI$ is altered by any set $\rm X$ of hypotheses that are used as discussed below. It is possible, however, that some of these $\rm R_i$ are $\rm N$ dependent. This means that various natural laws or processes can be incorporated within some of the n-ary relations, where $n > 1.$ It can be effectively decided when an $\rm x \in \rm L$ is a member of $\rm A\cup \rm N$ or a member of any of the fixed 1-ary relations.  Further,  for any finite $\rm B \subset \rm L$ and an $(j +1)$-ary $\rm R_{\rm i} \in {\bf RI},\ \rm j\geq 1$ and any $\rm f \in \rm R_{\rm i},$ it is always assumed that it can be effectively decided whether the k-th coordinate value $\rm f(\rm k) \in \rm B,\ \rm k= 1,\ldots,\rm j.$ It is always assumed that a mental or equivalent activity called {\it deduction} from a set of hypotheses can be represented by a finite (partial) sequence of numbered (in order) steps $\rm b_1,\ldots,\rm b_{\rm m}$  with the final step $\rm b_{\rm m}$ the conclusion of the deduction. All of these steps are considered as represented by objects from the language $\rm L.$ Any such representation is composed either of the zero step,  indicating that there are no steps in the representation,  or one or more steps with the last numbered step being some $\rm m >0$. In this inductive step-by-step construction,  a basic rule used to construct this representation is the {\it insertion} rule. If the construction is at the step number $\rm m \geq 0,$ then the insertion rule,  {\bf I},  is the ``insertion of an hypothesis from $\rm X \subset {\rm L},$ or insertion of a member from the set $\rm A\cup \rm N,$ or the insertion of any member from any 1-ary relation,  and denoting this insertion by the next step number.'' If the construction is at the step number $\rm m > 0,$ then the {\it rules of inference},  {\bf RI},  are used to allow for an insertion of a member from $\rm L$ as a step number $\rm m+1,$ in the following manner. For any 
$(\rm j+1)$-ary $\rm R_{\rm i} \in {\bf RI},$ $\rm 1\leq j,$ and any $\rm f \in \rm R_{\rm i},$ if $\rm f(\rm k) \in \{\rm b_1,\ldots, \rm b_{\rm m}\},\ \rm k=1,\ldots,\rm j,$ then $\rm f(\rm j+1)$ can be inserted as a step number $\rm m+1.$ Note, in particular, how specific ``choices'' are an essential part of the process here termed as deduction. \par\smallskip

For this paper, note the possible existence of special binary relations {\bf J} that may be members of {\bf RI}. These relations are identity styled relations in that the first coordinate and second coordinates are identical. In scientific theory building, these are used to indicate that a particular set of natural laws or processes does not alter a particular premise that describes a natural-system characteristic. The statement represented by this premise remains part of the final conclusion. Scientifically this can be a significant fact. These {\bf J} relations are significant for the extended realism relation. The deduction is constructed only from the rule of insertion or the rules of inference as described in this and the previous paragraph. \par\smallskip 
It is not difficult to show that if you apply these procedures to obtain the final step as your deduction,  then these procedures are modeled by a finitary consequence operator. For the language $\rm L,$ a set of pre-axioms $\rm A\cup \rm N,$ a set {\bf RI} and any $\rm X \subset {\rm L},$ define the set map $\rm C_N,$ by letting $\rm C_N(X)$ be the set of {\bf all} members of $\rm L$ that can be obtained from $\rm X$ by ``deduction.'' Clearly,  by insertion
$\rm X \subset  C_N(X).$ Since $\rm C_N(X) \subset {L},$ then consider the statement $\rm C_N(C_N(X))$. Since no member of the set $\bf RI$ is altered by introducing a different set of hypotheses such as $\rm C_N(X),$ then this composition is defined. Let $\rm x \in C_N(C_N(X)).$  By definition,  $\rm x$ is the final step in a finite list $\{\rm b_{\rm i}\}$ of members from $\rm L.$ The steps in this finite ``deduction'' from which $\rm x \in L$ is obtained are the $\bf I$ steps, where added to these insertions are only members from $\rm C_N(X),$ while the {\bf RI} steps, as defined above, are fixed. \par
Suppose that $\rm b_{i}\in C_N(X)$ is any of these additional insertions. Simply construct a new finite sequence of steps by substituting for each such $\rm b_{\rm i}$ the finite sequence of steps from which  $\rm b_{\rm i}$ is the final step in deducing that $\rm b_{i}\in C_N(X)$. The resulting finite collections of steps are then renumbered. The final step in this new finite deduction is $\rm x.$ Since the reasons for all of the steps is either the original {\bf I} or {\bf RI}, and {\bf RI} contains predetermined n-ary relations that are not dependent upon any deduction, then the finite sequence obtained in this manner is a deduction for a member of $\rm C_N(X)$. Hence,  $\rm x \in C_N(X).$ Consequently,  $\rm C_N(C_N(X)) = C_N(X).$ The finitary requirement is obvious since there are only a finite number of steps in any deduction. Note that $\rm C_N(\emptyset) \supset B,$ where $\rm B$ is the set of all non-hypothesis $\rm x \in L$ such that $\rm x$ is a step obtained only by the rule $\bf I.$  Throughout the remainder of this paper,  it is assumed that all ``deductions'' follow these procedures and the corresponding consequence operator is defined as above. [Note: There are other methods to define the {\bf RI}. This is especially the case for finite languages. Any description for deduction that leads to an argument similar to the one presented above would, of course, verify that the corresponding defined set map is a consequence operator.]\par\medskip
\noindent{\bf 4. Axiomatic intrinsic natural laws or processes.}\par\medskip

For ``scientific deduction'' for a fixed science-community, i, consider as our rules of inference a collection ${\bf R_i} = {\bf RI}$  of {\bf all} of the ``rules of inference used by this specific scientific-community and allowed by their scientific method'' as they are applied to a  specified language $\Sigma_{\rm i},$ the language for ``their science.''  At present, this definition for $\bf R_i$ is rather vague. Hence, the existence of such a set $\bf R_i,$ the rules of inference for a science-community, is an assumption. Of course, as $\Sigma_{\rm i}$ changes, so might the $\bf R_i$ be altered. The ${\bf R_i}$ can also change for other valid reasons. From this, a specific ``science'' consequence operator $\rm S_{\rm N_{\rm i}}$ is generated for each set of pre-axioms $\rm A_{\rm i}\cup \rm N_{\rm i},$ where $\rm A_i$ are the basic logical axioms and $\rm N_i$ the natural laws or processes. For proper application,  the science consequence operator is applied to specific natural-systems, not those generally described. Thus $\rm S_{\rm N_{\rm i}}$ has physical meaning only when $\rm S_{\rm N_{\rm i}}$ is applied to an $\rm X$ where every member of $\rm X$ and $\rm S_{\rm N_{\rm i}}(\rm X)$ is a ``tagged'' statement
that identifies a specific natural-system (Herrmann, 1999). In all that follows, a particular $\rm U_{\rm i} \subset\Sigma_{\rm i}$ of natural laws or processes is considered as accepted by a science-community at this particular moment of time and they are stated using the language $\Sigma_{\rm i}.$  \par\smallskip 
 
The axiomatic consequence operator $\rm S_{\rm N_{\rm i}}\colon \power {\Sigma_{\rm i}} \to \power {\Sigma_{\rm i}},$ where $\rm S_{\rm N_{\rm i}}(\emptyset) \supset (\rm A_{\rm i}\cup \rm N_{\rm i}),$ can be reduced,  formally,  to an axiomless consequence operator on the language $\Sigma_{\rm i} - \rm S_{\rm N_{\rm i}}(\rm A_{\rm i}\cup \rm N_{\rm i})$ as shown by Tarski (1930, p. 67). In this paper, this single process is termed {\it relativization.} Let $\rm V = \{\rm A_{\rm i},  \rm N_{\rm i}.\}$ For each $\rm X \subset \Sigma_{\rm i} - \rm S_{\rm N_{\rm i}}(\rm A_{\rm i}\cup \rm N_{\rm i}),$ let $\rm S^{\rm V}_{\rm N_{\rm i}}(\rm X) =(\Sigma_{\rm i}- \rm S_{\rm N_{\rm i}}(\rm A_{\rm i}\cup \rm N_{\rm i}))\cap \rm S_{\rm N_{\rm i}}(\rm X).$ For this $\rm S_{\rm N_{\rm i}},$ the operator $\rm S^{\rm V}_{\rm N_{\rm i}}$ is a consequence operator on $\Sigma_{\rm i} - \rm S_{\rm N_{\rm i}}(\rm A_{\rm i}\cup \rm N_{\rm i})$ and has the property that $\rm S^{\rm V}_{\rm N_{\rm i}}(\emptyset) = \emptyset.$ Thus, using $\rm S_{\rm N_{\rm i}}(\rm A_{\rm i}\cup \rm N_{\rm i})$ as a set of axioms,  logical and physical,  $\rm S^{\rm V}_{\rm N_{\rm i}}$ behaves as if it is axiomless,  where the explicit natural laws or processes $\rm N_{\rm i}$ behave as if they are implicit. Since,  in general,  $\rm S_{\rm N_{\rm i}}(\rm A_{\rm i}\cup \rm N_{\rm i}) \subset \rm S_{\rm N_{\rm i}}(\rm X),$ the only consequences that are not but specific deductions from the pre-axioms $\rm A_i \cup N_i$ are members of $\rm S_{\rm N_{\rm i}}(\rm X) - \rm S_{\rm N_{\rm i}}(\rm A_{\rm i}\cup \rm N_{\rm i}),$ where the explicit $\rm X$ should not include members of $\rm S_{\rm N_{\rm i}}(\rm A_{\rm i}\cup \rm N_{\rm i})$.  
Physically,  $\rm S^{\rm V}_{\rm N_{\rm i}}$ is the exact operator that,  using implicitly such axioms as $\rm S_{N_{ i}} (A_{i} \cup N_{i}),$ characterizes the coalescing of a given fundamental collection of named and tagged objects in $\rm X$ and that creates a different natural-system or that alters natural-system behavior. The use of axiomless consequence operators is an important simplification.\par\smallskip 
 Applying the above to an entire family of science-communities, there is for an arbitrary science-community, i, a nonempty sequentially represented collection $\rm V_i = \{A_i,  \{N_{ij}\mid j \in \nat\}\}$ such that for any $\rm N_{ij} \in {V_i},$ the set map $\rm S^{V_i}_{N_{ij}}$ defined for each $\rm X \subset (\Sigma_i- (\bigcup \{S_{N_{ij}}(A_i \cup N_j)\mid j \in \nat\})) = \Lambda_i$ by $\rm S^{V_i}_{N_{ij}}(X) =\Lambda _i\cap S_{N_{ij}}(X)$ is a consequence operator defined on $\rm \Lambda_i.$ (The set $\nat$ is the natural numbers not including 0.) The family $\rm V_i$ may or may not be finite. In many cases, it is denumerably since to apply   $\rm S^{V_i}_{N_{ij}}$ to a specifically tagged description $\rm X$ certain parameters within the appropriate set of natural laws or processes must be specified so as to correspond to the specific $\rm X.$ Assume that the applicable set of natural laws or process $\{\rm N_{ij}\}$ is  the range of a sequence. This will not affect the conclusions since this yields that $\rm V_i$ can be finite or denumerable. Note that for some of the $\rm N_{nm}$ and some tagged $\rm X \subset \Lambda_i$ to which the $\rm N_{nm}$ either does not apply or does not alter, that $\rm S^{V_n}_{N_{nm}}(X) = X.$ For logical consistency, 
it is significant if there exists some type of unifying consequence operator that will unify the separate theories not only applied by a specific science-community (i), but within all of science. \par\medskip
\noindent {\bf 5. The extended realism relation}\par\medskip
In previous investigations, the realism relation was defined for a given consequence operator $\rm C_N$ to be, for each collection of words $\rm X$ from our language {\rm L}, the set $\rm C(X) - X,$ which, using relativization, generates a consequence operator defined on $\rm L - X.$ In actual practice, this is a poor general definition since it removes from the conclusions those members of $\rm X$ that may simply not be altered by $\rm N$ and remain a characteristic of the conclusion, while other natural-system characteristics may have been altered. Under our new definition for the rules of inference this possibility can be included. The steps in any ``deduction'' for a conclusions $\rm x \in C_N(X)$ are finite in number. Thus using the aspect of human intelligence called (finite) choice, it is always effectively possible to determine whether a {\bf J} type rule of inference is used to obtain $\rm x,$ where $\rm x \in X.$ The extended realism relation is applied after application of such an $\rm C_N.$\par\smallskip
To obtain a possibly new set of premises for application of the appropriate  consequence operator, the {\it extended realism relation} is an operator that removes from $\rm C_N(X)$, through application finite human choice, those members of $\rm X$ that are not obtained from the rules of inference. Let $\rm R^{C_N}_X \subset L$ represent those members so removed. Notice that by relativization to the set $\rm L - R^{C_N}_X$, $\rm C_N(X)$ generates a consequence operator now defined on $\power {\rm L - R^{C_N}_X}$. If each member of a set of consequence operators is being applied simultaneously, then the language for the next application is the intersection of all of the individual $\rm L - R^{C_N}_X.$ In this paper, consequence operators are viewed as being applied to a specific and fixed subset of $\power {\rm L}$ at a particular moment, say in cosmic time. This extended realism relation, the relativization process, is applied only after such an application, it is not considered as part of the original requirements for the generating consequence operator and no additional notation is applied that indicates the relativized operator is required for the next application.  \par\medskip 

\noindent {\bf 6. A hyperfinite unification for physical theories.}\par\medskip
Although all that follows can be applied to arbitrary science-communities, for notational convenience, consider but one science-community. Thus assume that there is one language for science $\Sigma$ and one sequentially represented countable family of natural laws or processes and logical axioms $\rm A_j \cup  N_j$ as well as one family of sequentially represented rules of inference $\bf R_j$ that generate each specific theory. Let sequentially represented $\rm V = \{A_j \cup N_j\mid \ j \in \nat\}.$ This yields the sequentially represented countable set of physical theories $\rm \{S_{N_{i}}\mid j \in \nat \}$ and the countable set $\rm \{S^V_{N_{j}}\mid j \in \nat \}$ of intrinsic sequentially represented consequence operators defined on $\rm \Sigma- (\bigcup \{S_{N_{j}}(A_j \cup N_j)\mid j \in \nat\}) = \Lambda$. The following theorem and corollary do not depend upon each member of $\rm \{S_{N_j}\mid j \in \nat\}$ being declared as a ``correct'' physical theory.\par\smallskip
Our interest is in the non-trivial application of, at the least, one of these theories to members of $\power {{ \Lambda}}.$\par\medskip
{\bf Definition 6.1.} A nonempty $\rm X \in \power {{\Lambda}}$ is called a {\it significant} member of $\power {\Lambda}$ if there exists some $\rm i \in \nat$ such that $\rm X \not= S^V_{N_i}(X).$ Let nonempty $\rm M \subset \power {\Lambda}$ be the set of all significant members of $\power {\Lambda}.$\par\medskip
 {\bf Definition 6.2.} Suppose you have a nonempty finite set ${\cal C}=\{\rm C_1,\ldots,C_m\}$ of general consequence operators, each defined on a language $\rm L_i,\  1\leq i \leq m.$ 
Define the operator $\rm \Pi C_m$ as follows: for any $\rm X \subset L_1 \times \cdots \times L_m$, using the projection $\rm pr_i,\ 1\leq i\leq m$, define $\rm \Pi C_m(X) = C_1(pr_1(X))\times \cdots \times C_m(pr_m(X)).$ \par\medskip
(Note: For $\rm X \subset L_1 \times \cdots \times L_m,$ the usual definitions for the projection map yields that if $\rm X \not=\emptyset,$ then for each $\rm i,$ $\rm pr_i(X) \not= \emptyset$. The converse also holds. 
For the case that $\rm X = \emptyset,$ since no $\rm L_i$ is empty, choose for each projection the only function that exists with empty domain and nonempty codomain, the empty function. Each of these projections has an empty range (Dugundji, 1966, p. 11). )\par\medskip
{\bf Theorem 6.3.} {\it The operator $\rm \Pi C_m$ defined on the subsets of ${\rm L_1 \times \cdots \times L_m}$ is a general consequence operator and if, at least, one member of $\cal C$ is axiomless, then $\rm \Pi C_m$ is axiomless. If each member of $\cal C$  is finitary and axiomless, then $\rm \Pi C_m$ is finitary. }\par\medskip
Proof. (a)  Let $\rm X \subset  {\rm L_1 \times \cdots \times L_m}$. Then for each $\rm i,\ 1\leq i \leq m,\ pr_i(X) \subset C_i(pr_i(X)) \subset L_i.$ But, $\rm X \subset pr_1(X) \times \cdots \times pr_m(X) \subset C_1(pr_1(X)) \times \cdots \times C_m(pr_m(X))\subset {\rm L_1 \times \cdots \times L_m}.$ Suppose that $\rm X \not=\emptyset.$ Then $\rm \emptyset\not= \Pi C_m(X) = C_1(pr_1(X)) \times \cdots \times C_m(pr_m(X)) \subset {\rm L_1 \times \cdots \times L_m}$. Hence, $\rm \emptyset \not= pr_i(\Pi C_m(X)) = C_i(pr_i(X)),\ 1 \leq i \leq m$ implies that $\rm C_i(pr_i(\Pi C_m(X))) = C_i(C_i(pr_i(X))) = C_i(pr_i(X)),\ 1 \leq i \leq m.$ Hence, $\rm \Pi C_m(\Pi C_m(X)) = \Pi C_m(X).$ Let $\rm X = \emptyset$  and no member of $\cal C$ is axiomless. Then each $\rm pr_i(X) =\emptyset.$ But, each $\rm C_i(pr_i(X)) \not= \emptyset$ implies that $\rm \Pi C_m(X) \not=\emptyset.$ By the previous method, it follows, in this case, that $\rm \Pi C_m(\Pi C_m(X)) = \Pi C_m(X).$ Now suppose that there is some $\rm j$ such that $\rm C_j$ is axiomless. Hence, $\rm C_j(pr_j(X)) = \emptyset$ implies that $\rm \Pi C_m(X)=C_1(pr_1(X)) \times \cdots \times C_m(pr_m(X)) = \emptyset,$ which implies that $\rm C_j(pr_j(\Pi C_m(X))) = \emptyset.$ Consequently, $\rm C_1(pr_1(\Pi C_m(X))) \times \cdots \times \rm C_m(pr_m(\Pi C_m(X))) = \emptyset.$ Thus, $\rm \Pi C_m (\Pi C_m (X)) = \emptyset$ and axiom (1) holds. Also in the case where at least one member of $\cal C$ is axiomless, then $\rm \Pi C_m$ is axiomless.
\par\smallskip 
(b) Let $\rm X \subset Y \subset  {\rm L_1 \times \cdots L_m}.$ 
For each $\rm i,\ 1 \leq 1\leq m,\ \ pr_i(X) \subset pr_i(Y)$, whether $\rm pr_i(X)$ is the empty set or not the empty set. Hence, $\rm C_i(pr_i(X)) \subset C_i(pr_i(Y).$ Therefore, $\rm \Pi C_m(X) = C_1(pr_1(X))\times \cdots\times C_m(pr_m(X))\subset C_1(pr_1(Y))\times \cdots\times C_m(pr_m(Y))= \Pi C_m(Y)$ and axiom (2) holds. Thus, $\rm \Pi C_m$ is, at least, a general consequence operator. \par\smallskip
(c) Assume that each member of $\cal C$ is finitary and axiomless and let $\rm x \in \Pi C_m(X)$ where, since  $\rm \Pi C_m$ is axiomless, $\rm X$ is nonempty. Then for each $\rm i,\ pr_i(x) \in C_i(pr_i(X)).$ Since each $\rm C_i$ is finitary and axiomless, then there is some nonempty finite $\rm F_i \subset pr_i(X)$ such that $\rm pr_i(x) \in C_i(F_i) \subset C_i(pr_i(X)).$ Hence, nonempty and finite $\rm F = F_1 \times \cdots \times F_m \subset pr_1(X) \times \cdots \times pr_m(X).$ Then for each $\rm i,\ pr_i(F) = F_i$ implies that finite $\rm F = F_1 \times \cdots \times F_m = pr_1(F)\times \cdots \times pr_m(F) \subset pr_1(X) \times \cdots \times pr_m(X).$ From axiom (2), $\rm x \in \Pi C_m(F) = C_1(pr_1(F)) \times \cdots \times C_m(pr_m(F))\subset \Pi C_m(pr_1(X) \times \cdots\times pr_m(X)) = C_1(pr_1(X)) \times \cdots \times C_m(pr_m(X)) = \Pi C_m(X).$ This completes the proof.\qed\par\medskip 

In what follows, consider all of the previously defined notions but only with respect to this informal $\rm V$ and the language ${\Lambda}.$ (Although, the consequence operators are being restricted to a special collection that is of interest to various science-communities, Theorem 6.4 will hold, with obvious modifications, for any sequentially represented set of consequence operators.) Although strictly not necessary, in order for the following to correlate with the results in Herrmann (2001a), embed all of these informal results into the formal superstructure ${\cal M} = \langle {\cal R}, \in ,= \rangle$ as done in Herrmann (1993, p. 70) where $\cal R$ is isomorphic to the real numbers. Further, consider the structure $\Hyper {\cal M} = \langle\Hyper {\cal R}, \in,=\rangle$ a nonstandard and elementary extension of $\cal M$ that is a $2^{\vert {\cal M} \vert}$-saturated enlargement ($\vert \cdot \vert$ denotes cardinality). Finally, consider the superstructure ${\cal Y},$ the {\it Extended Grundlegend Structure}. Note that a structure based upon the natural numbers appears adequate for our analysis since this investigation is only concerned with members of a denumerable language. However, $\cal Y$ is used so that the results here can be directly related to those in Herrmann (2001a). par\smallskip
A unifying consequence operator approach seems at first to be rather obvious. In actual physical practice, a set of physical theories is applied to a specific $\rm W \subset  \Lambda.$  The result is $\rm C(W) = \bigcup\{S^V_{N_j}(W)\mid j \in \nat \}$ and is most certainly a unification for all of the physical science theories $\rm S_{N_j}.$ But, define the consequence operator $\rm B$ on $\power {\rm \{a,b,c,d,e\}}$ by two relations $\{\rm (a,b,c),\ (d,e)\}.$  Then $\rm B(\{a,b\}) = \{a,b,c\}$ and $\rm B(\{d\}) = \{d,e\}.$ Define $\rm R$ on $\power {\rm \{a,b,c,d,e\}}$ by one relation $\{\rm (a,b,d)\}.$ Hence, $\rm R(\{a,b\}) = \{a,b,d\}.$ The union operation $\rm K$ is defined for each $\rm X \in \power{\rm \{a,b,c,d,e\}}$ by $\rm K(X) = B(X) \cup R(X).$ However, $\rm K(\{a,b\}) = \{a,b,c,d\}$ and $\rm K(K(\{a,b\})) = \{a,b,c,d,e\}$ and axiom (1) does not hold. The union operation, even in the simplest sense, is not determine a consequence operator. How can this actual physical practice be considered as a rational process? This is done by adjoining the choice aspect of intelligence to the process. Applying this additional step, there is a standard consequence operator styled  unification and more than one ultralogic styled unification. \par\smallskip
The following ultralogic styled unification is consistent with the modeling of the ultralogic generation of probabilistic behavior (Herrmann, 1999, 2001a) and minimal in the hyperfinite sense. In ultralogic theory, attempts are made to stay within the bounds of the ``finite'' or ``hyperfinite.'' In all known cases, the set of hypotheses selected from $\rm M$ is a finite set. For such selections of finite sets of significant hypotheses, the standard consequence operator $\rm P$ used to establish Theorem 6.4 is a finitary consequence operator. It is a {\it practical} consequence operator.
 \par\medskip
\vfil\eject

{\bf Theorem 6.4.} {\it Given the language $ {\Lambda}$ and the sequentially represented set of consequence operators $\rm \{S^V_{N_j}\mid j \in \nat \}.$ \par\smallskip
{\rm (i)} There exists an injection ${\cal S}$ on nonempty $\rm M,$ the set of all significant subsets of ${\Lambda},$ into $\Hyper {(\bf {\cal C}({\bf \Lambda}))}$ such that for each $\rm W  \in M,$ ${\cal S}_{\rm W}$ is a   hyperfinite consequence operator, an ultralogic, such that $\bigcup\{{\bf S}^{\bf V}_{{\bf N}_j}({\bf W})\mid j \in \nat \}\subset \bigcup\{\Hyper {\bf S}^{\bf V}_{{\bf N}_j}(\Hyper {\bf W})\mid j \in \nat \}= \bigcup\{\Hyper {({\bf S}^{\bf V}_{{\bf N}_j}({\bf W}))}\mid j \in \nat \}\subset {\cal S}_{\rm W}(\Hyper {\bf W})$ and $\bigcup\{{{{\bf S}^{\bf V}_{{\bf N}_j}({\bf W})}}\mid j \in \nat \}=  {\cal S}_{\rm W} (\Hyper {\bf W} ) \cap {\bf \Lambda}.$\par\smallskip
{\rm (ii)} If $\rm \emptyset \not=  \{W_1,\ldots,W_m\} \subset M,$ then there exists a hyperfinite consequence operator $\Pi$ defined on internal subsets of the product set $\hyper {\bf \Lambda}^m$ such that for each $i = 1,\ldots m,\ {\cal S}_{\rm W_i} (\Hyper {\bf W_i}) = \Hyper {pr_i}(\Pi (\Hyper {\bf W_1} \times \cdots \times \Hyper {\bf W_m}))$. If each $\rm W_i,\ 1 \leq i \leq m,$ is finitary, then $\Pi$ is hyperfinitary.
\par\smallskip
{\rm (iii)} For each $\rm W, \ A \in M,$ if $\rm W \subset A$ then
${\cal S}_{\rm W}(\Hyper {\bf W}) \subset {\cal S}_{\rm W}(\Hyper {\bf A})\subset {\cal S}_{\rm A}(\Hyper {\bf A}).$}\par\medskip 
Proof. (i) In Herrmann (1987, p. 4), special sets of consequence operators are defined. For this application and for a given $\rm X \in  M$, the set is $\rm H_X = \{P(Y,  X )\mid Y \subset  \Lambda \}.$ Each of the consequence operators in $\rm H_X$ is defined as follows: for each $\rm Z \subset \Lambda,\  P(Y,X)(Z) = Y \cup Z,$ if $\rm X \subset Z$; and $\rm P(Y,X)(Z)=Z$ otherwise.  It is shown in Herrmann (1987) that $\rm P$ is a general consequence operator; if $\rm X$ is nonempty, then $\rm P$ is axiomless; if $\rm X$ is finite, then $\rm P$ is finitary.  Let $\rm \emptyset\not= {\cal A}\subset \power {\Lambda}.$  Suppose that $\rm X \subset Z.$ Then $\rm P(\cup {\cal A},X)(Z)= (\cup {\cal A})\cup Z = \bigcup \{A\cup Z\mid A \in {\cal A}\} = \bigcup \{P(A,X)(Z)\mid A \in {\cal A}\}.$ Now suppose that $\rm X \not\subset Z.$ Then  $\rm P(\cup {\cal A},X)(Z)= Z = \bigcup \{P(A,X)(Z)\mid A \in {\cal A}\}\in H_X.$   Thus $\rm H_X$ is closed under the arbitrary union operation.   \par\smallskip
Consider the entire set of intrinsic consequence operators $\rm \{S^V_{N_j}\mid j \in \nat \}.$ Define by induction, with respect to the sequentially represented $\rm \{S^V_{N_{i}}\mid j \in \nat \},$
 $\rm P_1 = P(S^V_{N_1}(X),X),\ P_2= P(S^V_{N_1}(X) \cup S^V_{N_2}(X), X ),\ldots, P_n= P(S^V_{N_1}(X) \cup \cdots \cup S^V_{N_n}(X), X ).$ From this definition, it follows that for any $\rm n \in \nat$ the equation (*) $\rm P_n(X)= S^V_{N_1}(X) \cup \cdots \cup S^V_{N_n}(X)$ holds for each $\rm X \subset  \Lambda$. The $\rm P_n,$ therefore, unify the finite partial sequences of $\rm \{S^V_{N_j}\mid j \in \nat \}.$  This restriction to but finite unions is the aspect that allows for a type of minimal ultralogic to be generated. \par\smallskip
All of the above is now embedded into $\cal M$ and then considered as embedded into the superstructure $\cal Y.$  Since $\rm \{S^V_{N_{i}}\}$ is sequentially represented, there is a fixed sequence $g$ such that $g(i) = {\bf S^V_{N_i}},\  g[\nat] =  \{{\bf S}^{\bf V}_{{\bf N}_j}\mid j \in \nat \}$ and $g(i)({\bf X}) = {{\bf S^V_{N_i}}}({\bf X}).$ Hence for arbitrary $\rm X \subset \Lambda$, utilizing  $g$, the above inductive definition yields a sequence $f_{\bf X} \colon \nat \to {\bf H_X}$ such that $f_{\bf X}(j) = {\bf P_j}$ and $f_{\bf X}(j)({\bf X}) = {\bf P_j}({\bf X})$ and, as embedded into $\cal M$, equation ({\bf *}) holds.\par\smallskip
Let $\rm X \subset \Lambda.$ Then the following sentence holds in $\cal M.$
$$\forall x\forall i((x \in {\bf \Lambda}) \land (i \in \nat) \to  (x \in  f_{\bf X}(i)( {\bf X} ) \iff $$ $$ \exists j ((j \in \nat) \land (1\leq j \leq i)\land (x \in g(j)({\bf X}))))).\eqno (6.1)$$\smallskip
\noindent By *-transfer, the sentence
$$\forall x\forall i((x \in 
\hyper{{\bf \Lambda}}) \land (i \in \hypernat) \to  (x \in  \Hyper {(f_{\bf X}(i)( {\bf X} ))} \iff $$ $$ \exists j ((j \in \hypernat) \land (1\leq j \leq i)\land (x \in \Hyper {(g(j)({\bf X}))}))))\eqno (6.2)$$\smallskip
\noindent holds in $\Hyper {\cal M}.$ Due to our method of embedding and identification, sentence (6.2) can be re-expressed as
$$\forall x\forall i((x \in 
\hyper{{\bf \Lambda}}) \land (i \in \hypernat) \to  (x \in  \hyper {f_{\bf X}(i)}( \Hyper {{\bf X}}) \iff $$ $$ \exists j ((j \in \hypernat) \land (1\leq j \leq i)\land (x \in \hyper {g(j)}(\Hyper {{\bf X}})))))\eqno (6.3)$$\smallskip
\noindent Next consider $\hyper {f_{\bf X}} \colon \hypernat \to \Hyper {\bf H_X}$ and any $\lambda \in \hypernat - \nat.$ Then hyperfinite $\hyper {f_{\bf X}}(\lambda) \in \Hyper {\bf H_X}$ is a nonstandard consequence operator, an ultralogic, that is hyperfinite in the sense that it hyperfinitely generates each of the $\Hyper {\bf P_j}$ and it satisfies  statement (6.3). Hence, arbitrary $j \in \nat$
and $w \in \hyper {g}(j)(\Hyper {\bf X}) = \Hyper {\bf S^V_{N_j}}(\Hyper {\bf X})=\Hyper {({\bf S^V_{N_j}(X)})}\subset \hyper {{\bf \Lambda}}$ imply that $w \in \hyper {f_{\bf X}(\lambda)}( \Hyper {\bf X} )$ since $1 \leq j < \lambda.$ Observe that ${}^{\sigma} {({\bf S^V_{N_j}(X)})}\subset \Hyper {({\bf S^V_{N_j}(X)})}.$ However, under our special method for embedding ${}^{\sigma} ({\bf S^V_{N_j}(X)}) = {\bf S^V_{N_j}(X)},$ for an arbitrary $\bf X \subset {\bf \Lambda}.$\par\smallskip

The final step is to vary the $\rm X \in M.$ It is first shown that for two distinct $\rm X,\ Y \in {\rm M}$ there is an $\rm m \in \nat$ such that 
$\rm P_m^X = P(S^V_{N_1}(X) \cup \cdots \cup S^V_{N_m}(X), X) \not= P_m^Y = P(S^V_{N_1}(Y) \cup \cdots \cup S^V_{N_m}(Y), Y).$ Since $\rm X,\ Y$ are nonempty, distinct and arbitrary, assume that $X \not\subset Y.$ Hence there is some $\rm i \in \nat$ and $\rm j \in \nat$ such that $\rm X \subset S^V_{N_i}(X) \not= X$ and  $\rm Y \subset S^V_{N_j}(Y) \not= Y.$
Consider some $\rm m \in \nat$ such that $\rm i,\ j \leq m.$ Let $\rm Y \subset X.$ Then 
$\rm P_m^X(X) = P(S^V_{N_1}(X) \cup \cdots \cup S^V_{N_m}(X), X)(X)= 
S^V_{N_1}(X) \cup \cdots \cup S^V_{N_m}(X) \not= X\subset P_m^X(X).$ But $\rm P_m^Y(X) = X \subset P_m^X(X)\not= X.$ Thus $\rm P_m^Y(X)\not=  P_m^X(X).$
Now further suppose that $\rm Y\not\subset X.$ Then there is some $\rm y\in Y$ such that $\rm \{y\} \not\subset X.$ Since $\rm X \subset X \cup \{y\}$ and $\rm Y \not\subset X \cup \{y\}$, then  $\rm P_m^X(X\cup \{y\}) =  
(S^V_{N_1}(X) \cup \cdots \cup S^V_{N_m}(X)) \cup \{y\}.$ Since $\rm Y\not\subset X \cup \{y\}$, then $\rm P_m^Y(X \cup \{y\})= X \cup \{y\}.$ Again since $\rm S^V_{N_1}(X) \cup \cdots \cup S^V_{N_m}(X) \not=X,$ then 
$\rm P_m^Y(X \cup \{y\}) \not= P_m^X(X \cup \{y\}).$ In like manner, if $\rm Y \not\subset X.$ Consequently, $\rm P_m^Y\not= P_m^X.$ Further, for any ($\dagger$) $\rm k \in \nat,\ m \leq k,\  P_k^Y\not=  P_k^X.$ Consider these results formally stated. Then  by *-transfer, for each distinct pair ${\bf X},\ {\bf Y} \in {\bf M}$ there exists some $m \in \hypernat$ such that $\hyper {f_{\bf X}}(m) \not= \hyper {f_{\bf Y}}(m).$ Thus for ${\bf X}, \ {\bf Y} \in {\bf M}, \  {\bf X} \not= {\bf Y},$  $A({\bf X}, {\bf Y}) =\{m 
\mid (m \in \hypernat)\land \hyper {f_{\bf X}}(m) \not= \hyper {f_{\bf Y}}(m)\}$ is nonempty.  
The Axiom of Choice for the  general set theory (Herrmann, 1993, p. 2) used to construct our $\cal Y$ is now applied. Hence, there exists a set $B,$ within our structure, containing one and only member from each $A({\bf X}, {\bf Y}).$\par\smallskip
The internal binary relation $\{(x,y)\mid (x \in \hypernat)\land (y \in \hypernat)\land  (x \leq y)\}$ is, from *-transfer of $\nat$ properties, a concurrent relation with respect to the range $\hypernat$. Since $\Hyper {\cal M}$ is  a $2^{\vert {\cal M} \vert}$-saturated enlargement and $\vert B \vert < 2^{\vert {\cal M} \vert}$, there is some $\lambda \in \hypernat$ such that for each $i \in B,\ i \leq \lambda$. Considering this $\lambda$ as fixed, then by *-transfer of $(\dagger)$, it follows that for any distinct $\rm X ,\ Y \in M$ $\hyper {f_{\bf X}}(\lambda) \not= \hyper {f_{\bf Y}}(\lambda).$ Since $\rm M$ is injectively mapped onto $\bf M$, there exists an injection $\cal S$ on the set $\rm M$ such that each $\rm W \in M$, $ {\cal S}_{\rm W}= \hyper {f_{\bf W}}(\lambda) \in  \Hyper {({\bf {\cal C}(\Lambda)})}.$   Considering the general properties for such an $\hyper {f_{\bf W}}(\lambda)$ as discussed above, it follows that $\bigcup\{{{{\bf S}^{\bf V}_{{\bf N}_j}({\bf W})}}\mid j \in \nat \}\subset {\cal S}_{\rm W} (\Hyper {\bf W} ) \cap {\bf \Lambda}.$\par\smallskip
Now assume that standard ${\bf a} \in {\cal S}_{\rm W} (\Hyper {\bf W} )- \bigcup\{{{{\bf S}^{\bf V}_{{\bf N}_j}({\bf W})}}\mid j \in \nat \}.$ (For our identification and embedding, $\hyper {\bf a} = {\bf a}.$) Then the following sentence
$$ \forall x \forall i( (x \in  {\bf \Lambda}) \land (i \in \nat) \land x \in g(i)({\bf W}) \to x \not= {\bf a}) \eqno (6.4)$$
holds in $\cal M$ and, hence,
$$ \forall x \forall i( (x \in \hyper{ {\bf \Lambda}}) \land (i \in \hypernat) \land x \in \hyper {g(i)}(\Hyper {\bf W}) \to x \not= {\bf a}) \eqno (6.5)$$
holds in $\Hyper {\cal M}.$ But since ${\bf a} \in \hyper {f_{\bf W}}(\lambda)(\Hyper {\bf W}),$ then statement (6.5) contradicts statement (6.3) and this completes the proof of (i).\par\smallskip
(ii) Note that for each $\rm W \in M,$ $\Hyper { f_{\bf W}(\lambda)}$ is axiomless by *-transfer since every member of ${\Hyper {\bf H_W}}$ is axiomless. Consider nonempty $\rm \{W_1, \ldots,W_m\} \subset M.$ Use definition 6.2 and define $\Pi$ on the internal $X \in \Hyper {(\power {{\bf \Lambda}^m})}= \Hyper {\power { \Hyper {\bf \Lambda}^m}}$ by $\Pi(X) =\hyper {f_{\bf W_1}}(\lambda)(\Hyper {pr_1}(X))\times \cdots \times \hyper {f_{\bf W_m}}(\lambda)(\Hyper {pr_m}(X)).$ It is easily seen by *-transfer of Theorem 6.2 that $\Pi$ is a hyperfinite ultralogic. Or, more directly, notice that the properties of the projection maps $\Hyper pr_i$ on internal sets are the same as the standard projection maps. Since each member of $\Hyper {\bf H_{W_i}}$ satisfies the *-transfer of axioms (1) and (2) on internal sets, then for $\hyper {f}_{\bf W_i}(\lambda)$ this leads to the exact same (a) (b) proofs as for Theorem 6.3. Since the finite Cartesian product of hyperfinite objects is hyperfinite, then $\Pi$ is hyperfinite. Therefore, $\Pi$ is a hyperfinite ultralogic and for each $i,\ 1\leq i \leq m, \ \Hyper {pr_i}(\Pi(\Hyper {\bf W_1} \times \cdots \times \Hyper {\bf W_m})) = \hyper {f_{\bf W_i}}(\lambda)(\Hyper {\bf W_i}) = {\cal S}_{\rm W_i}(\Hyper {\bf W_i}).$  Finally, if each $\rm W_i$ is finite, then each $\hyper {f_{\bf W_i}}(\lambda)$ satisfies the *-transfer of axiom (3). This means that in the place of the ``finite'' sets, hyperfinite sets are  utilized. Again, since the finite Cartesian product of hyperfinite sets is hyperfinite, then $\Pi$ is hyperfinitary.\par\smallskip
(iii) For each $\rm j \in \nat,\ W,\ A \in M,$ where $\rm W \subset A,$ it 
follows that $\rm P^W_j(W) \subset P^W_j(A) = S^V_{N_i}(W) \cup \cdots \cup S^V_{N_j}(W) \cup A.$ But, $\rm P^A_j(A) = S^V_{N_1}(A) \cup \cdots \cup S^V_{N_j}(A) \cup A = S^V_{N_1}(A) \cup \cdots \cup S^V_{N_j}(A).$ Since for  each $\rm i \in \nat,\ S^V_{N_i}(W) \subset S^V_{N_i}(A),$ then $\rm 
P^W_j(A) \subset P^A_j(A).$ Thus, after embedding, $f_{\bf W}(j)({\bf W}) \subset  f_{\bf W}(j)({\bf A}) \subset f_{\bf A}(j)({\bf A}).$ By *-transfer, this yields that ${\cal S}_{\rm W}(\Hyper {\bf W})=\hyper {f}_{\bf W}(\lambda)(\Hyper {\bf W}) \subset  {\cal S}_{\rm W}(\hyper {\bf A})=\hyper {f}_{\bf W}(j)(\hyper {\bf A}) \subset {\cal S}_{\rm A}(\hyper {\bf A})=\hyper {f}_{\bf A}(j)(\hyper {\bf A}).$ This completes the proof. \qed
{\bf Corollary 6.5.} {\it If $\rm \{S^V_{N_j} \mid j \in \nat\}$ represents all of the physical theories that describe natural world behavior, then the choice function and the last equation in part (i) of Theorem 6.4 and part (ii) correspond to an ultralogic unification for $\rm \{S^V_{N_j} \mid j \in \nat\}$}.\par\medskip 
{\bf Remark 6.6} Obviously, the results of Theorem 6.4 (i), (iii) and Corollary 6.5 (i) do not require that the standard theory consequence operators be axiomless. Further, in proofs such as that of Theorem 6.4, some of the results can be obtained without restricting the construction to members of $\rm H_X.$ Simply define, for $\rm X \subset \Lambda, \ P'_n(X)= S^V_{N_1}(X) \cup \cdots \cup S^V_{N_n}(X).$ One obtains a hyperfinite $\hyper {h}_{\bf W}(\lambda)$ from this definition. The use of the $\rm H_X$ objects is to further identify  $\hyper {f}_{\bf W}(\lambda)$ as an ultralogic that is also hyperfinite and of the type used in Herrmann (1999, 2001a) and, hence, allows for the modeling of certain aspects of intelligence. In general, a basic aspect of intelligence is to select a specific member of $\rm M$ to which to apply a specific physical theory while rejecting other members of $\rm M$ as so applicable. Due to the construction of physical theory consequence operators with the extended realism relation, this special aspect of intelligence is not usually modeled. But, this additional aspect of intelligence is modeled by the choice procedures in the proof of Theorem 6.4 and the specific selection of a member of $\rm H_X.$ \par\medskip  
Note that usually $\rm W$ is a finite set. Assuming this case, then again due to our method of embedding $\Hyper {{\bf W}} = {\bf W}.$ In statement (3), $\Hyper {g(i)} = \Hyper {{\bf  S^V_{N_i}}}.$ However, $\rm S^V_{N_i}$ has had removed all of the steps that usually yield an infinite collection of results when $\rm S_{N_i}$ is applied to $\rm W.$ Thus, in most cases, 
$\rm S_{N_i}(W)$ is a finite set. Hence, assuming these second finite case, it follows that ${\bf S^V_{N_j}(W)}= \Hyper {({\bf S^V_{N_j}(W)})}.$ However, each ${\cal S}_{\rm W}$ remains a nonstandard ultralogic since each ${\cal S}_{\rm W}$ is defined on the family of all internal subsets of $\hyper { {\bf \Lambda}}$ and the combined collection of all of the scientific theories implies that $\Lambda$ is denumerable. Of significance is that corollary 6.5 is technically falsifiable. The most likely falsifying entity would be the acceptance of a physical theory that does not use the rules of inference as setout in section 3. In particular, when different hypotheses are considered, the requirement that the rules of inference {\bf RI} cannot be altered.  \par\smallskip
 Such operators as ${\cal S}_{\rm W}$ can be interpreted in distinct ways. If they are interpreted in a physical-like sense, then they operate in a region called the {\it nonstandard physical world} (Herrmann, 1989), where $\rm W$ corresponds physically to the natural-system it describes. The restriction ${\cal S}_{\rm W}( {\Hyper {\bf W}} ) \cap { {\bf \Lambda}}$ then represents a natural world entity. As a second interpretation, $\cal S$ would represent an intrinsic process that appears to guide the development of our universe and tends to verify the Louis de Broglie statement. ``[T]he structure of the material universe has something in common with the laws that govern the workings of the human mind'' (March, 1963, p. 143). 
 \par\medskip
\noindent {\bf 7. A standard consequence operator unification.}\par\medskip
In practical science, each $\rm S^V_{N_j}$ is applied to a finite $\rm X \subset \Lambda.$  In Herrmann(2001b), Theorem 5.1 is the same as Theorem 6.4 part (i) in this paper. There is a major difference, however, in how the results are obtained. In Theorem 6.4, a considerably different standard consequence operator $\rm P$ is used. Since, in general, the union operation does not generate a consequence operator, any unifying consequence operator standard or nonstandard would have additional characteristics. These characteristics considered from the physical theory viewpoint might be rather undesirable. For example, the standard consequence operator $\rm C^X_m$ used in Theorem 5.1 (Herrmann, 2001b) when applied at any nonempty $\rm Y \subset X,$ where $\rm Y, X \in M$, has the property that $\rm C^X_m(Y) = C^X_m(X) = S^V_{N_1}(X) \cup, \cdots, \cup S^V_{N_m}(X).$ Although the choice function does not allow an application to objects distinct from $\rm X,$ the fact that $\rm C^X_m$ satisfies the consequence operator axioms does require such an application. Can such necessary secondary requirements be ignored based upon practical physical usage? From the standpoint of such notions as cosmic time, the answer is yes. What actually is assumed to occur is that such an unification is applied throughout the entire universe at each moment in cosmic time.\par\smallskip
The example given in section 6 that the union operation does not yield a consequence operator shows that the idempotent property fails. Although it is not a necessary requirement for the next result, under the assumption that all scientific theories must be consistent in combined form, then a practical union operation should share the same rationality as the individual $\rm S^V_N$ operators. \par\medskip
{\bf Theorem 7.1.} {\it Define for each $\rm X \in \power {\Lambda},$ the operator $\rm P^X_\infty =P(\bigcup\{S^V_{N_j}(X)\mid j \in \nat \},X)\in H_X.$\par\smallskip {\rm (i)} There is an injection $\rm S$ on $\rm M$ such that for each $\rm Z \in M$, $\rm S_Z = P^Z_\infty$. \par\smallskip 
{\rm (ii)} For each $\rm W \in M$, and $\rm S_W(W),$ the set $\rm S_W(A), \ A\not=W, \ A \in M,$ is consistent with the alterations in natural-system behavior modeled by $\rm S_W(W)$.  Further, if $\rm W \subset A,$ then
$\rm S_W(W) \subset S_W(A) \subset S_A(A).$ 
\par\smallskip 
{\rm (iii)} If $\rm \emptyset \not=  \{W_1,\ldots,W_m\} \subset M,$ then there exists a consequence operator $\Pi$ defined on $\power {{\Lambda}^m}$ such that for each $\rm i = 1,\ldots m,\ S_{\rm W_i}(W_i) =  pr_i(\Pi (\rm W_1 \times \cdots \times W_m))$. If each $\rm W_i,\ 1 \leq i \leq m,$ is finite, then $\Pi$ is finitary.
\par\smallskip
{\rm (iv)} For each $\rm W \in M,\ {\cal S}_{W}(\Hyper {\bf W}) \subset \Hyper {\bf S_W}(\Hyper {\bf W})$.}
\par\smallskip
Proof. (i) The proof in Theorem 6.4 that shows that for distinct $\rm W,\ Y \in M$
$\rm P^W_m \not= P^Y_m$ also holds for $\rm P^W_\infty$ and $\rm P^Y_\infty.$ Hence, simply defined a map $\rm S$ on $\rm M$ by $\rm S_W = P^W_\infty.$ This map is an injection. \par\smallskip
(ii) If $\rm W \in M,$ then $\rm P^W_\infty(W) =\bigcup\{S^V_{N_j}(W)\mid j \in \nat \}$ follows from the definition  of the $\rm P$ consequence operator. Now assume that $\rm A \in M,\ W \not= A.$ First, let $\rm W  \subset A.$ Thus $\rm W$ represents a subsystem of the natural-system $\rm A$ that is distinct from $\rm A.$ Then $\rm P^W_\infty(W)(A)= \bigcup\{S^V_{N_j}(W)\mid j \in \nat \} \cup A$. It is the adjoined $\rm A$ that allows for the idempotent portion of axiom (1). Is this adjoined $\rm A$ of any physical concern? Since $\rm W \subset A$ is a very general statement without any other characterizing requirements, it is not known whether $\rm A \subset \bigcup\{S^V_{N_j}(W)\mid j \in \nat \}.$ Thus the results can be interpreted as stating that when $\rm P^W_\infty$ is applied to a subsystem $\rm W \subset A$ the result contains those members $\rm A$ that are not altered by any of the $\rm S^V_{N_j}$ applied to $\rm W.$ Of course, one my also assume that at the same instant $\rm P^A_\infty$ is applied to $\rm A.$    
This yields that $\rm P^W_\infty(W) \subset P^A_\infty(A)$ which is exactly as one would expect. In the case that $\rm W\not\subset A,$ then $\rm P^W_\infty(A) = A$. This is consistent with the lack of a specific characterizing sharing relation between the $\rm W $ and $\rm A$ natural-systems. Of course, $\rm P^A_\infty(P^W_\infty(A))=  P^A_\infty(A) = \bigcup\{S^V_{N_j}(A)\mid j \in \nat \}.$ Thus neither of these results appears to have any significance physical inconsistencies. The statement that 
if $\rm W \subset A,$ then
$\rm S_W(W) \subset S_W(A) \subset S_A(A)$ is established in the same manner as part (iii) of Theorem 6.4.\par\smallskip
(iii) Theorem 6.3.\par\smallskip
(iv)  From Theorem 6.4, for each ${\rm W \in M,\ {\cal S}_W(\Hyper {\bf W})} = \hyper {f}_{\bf W}(\lambda)(\Hyper {\bf W}).$ From the definition of the sequence $f$ the following
$$\forall i\forall x((x \in {\bf \Lambda})\land (i \in \nat) \land (x \in f_{\bf W}(i)({\bf W})) \to x \in {\bf S_W(W)})\eqno (7.1)$$
holds in $\cal M$. Hence, by *-transfer
$$\forall i\forall x((x \in \hyper {\bf \Lambda})\land (i \in \hypernat)\land (x \in \hyper {f}_{\bf W}(i)(\Hyper {\bf W})) \to x \in \Hyper {{\bf S_W}}(\Hyper {\bf W})). \eqno (7.2)$$ 
However, letting $i = \lambda$ and noting that $\hyper {f}_{\bf W}(\lambda)(\Hyper {\bf W}) = {\cal S}_{\rm W}(\Hyper {\bf W}),$ then this completes the proof. \qed
{\bf Remark 7.2.} Expressions such as (6.4) and (7.1) are equivalent to expressions written in the required ``bounded" formalism since the standard superstructure is closed under basic set-theoretic operations. Further, from *-transfer, the basic results in Theorem 7.1 part (ii) also hold for $\hyper {\bf S}_{\bf W}(\Hyper {\bf W})$ and $\hyper {\bf S}_{\bf A}(\Hyper {\bf A})$ as well as internal arguments in general.\par\medskip
\noindent {\bf 8. Probability models.}\par\medskip
In Herrmann (1999, 2001a), it is shown that given a specific probability theory for a specific source or natural-system described by a single sentence $\{\rm G\}$ that predicts that an event $\rm E$ will occur with probability $p$ then there is an ultralogic $P_p$ that generates an exact sequence of such events the relative frequency of which will converge to $p.$ It is also shown that the patterns produced by the frequency functions for statistical distributions that model natural-system behavior are also the results of applications of ultralogics. Although the main results in these papers state as part of the hypothesis that $p$ is theory predicted, the results also hold if $p$ or the distribution is obtained from but empirical evidence. Theorem 2.1 in Herrmann (2001a) actually corresponds to Theorem 6.4. Notice that in Theorem 2.1 in Herrmann (2001a), that singleton type sets $\{\rm G\}$  are the only sets to which the basic consequence operator is applied. Thus, throughout Theorem 2.1 you can substitute for the $C$ operator, the operator $\rm P$ and the same results are obtained.   \par\smallskip
Are these results for probability models consistent with Theorem 6.4?
If probability models predict natural-system behavior, in any manner, then, in general, the natural laws or processes $\rm N$ that are assumed to lead to such behavior only include a statement that claims that the event sequences or distributions appear in the natural world to be ``randomly'' generated. It is precisely the results in Herrmann (1999, 2001a) that show that in the nonstandard physical world such behavior need not be randomly obtained but can be specifically generated by ultralogics. These results are thus consistent since the ultralogics obtained from Theorem 2 neither correspond to nor apply to any nonstandard extension of the notion of standard ``randomness.''\par\medskip  

\centerline{\bf References}\par\medskip
\id{D}ugundji, James. (1966), {\it Topology.} Boston: Allyn and Bacon.
\id{F}erris,  Timothy. (1977), {\it The Red Limit.}  New York: Bantam Books.\smallskip
\id{H}errmann, Robert A. (2001a), ``Ultralogics and probability models,'' {\it Int. J. of Math. Math. Sci.} (To appear).\smallskip
\id{H}errmann, Robert A. (2001b), ``An ultralogic unification for all physical theories,'' http://www.arXiv.org/abs/physics/0101009.\smallskip
\id{H}errmann, Robert A. (1999), ``The wondrous design and non-random character of `chance' events,'' http://www.arXiv.org/abs/physics/9903038\smallskip
\id{H}errmann, Robert A. (1993),  {\it The Theory of Ultralogics.} \hfil\break http://www.arXiv.org/abs/math.GM/9903081 and/9903082\smallskip

\id{H}errmann, Robert A. (1989), ``Fractals and ultrasmooth microeffects.'' {\it Journal of Mathematical Physics} 30(4):805-808. \smallskip

\id{H}errmann, Robert A. (1987), ``Nonstandard consequence operators''. {\it Kobe Journal of  Mathematics} 4:1-14. http://www.arXiv.org/abs/math.LO/9911204\smallskip

\id{M}arch, Arthur and Ira M. Freeman. (1963), {\it The New World of Physics}. New York: Vintage Books.\smallskip

\id{T}arski,  Alfred. (1956), {\it Logic, Semantics, Metamathematics; papers from 1923 - 1938},  Translated by J. H. Woodger$.$ Oxford: Clarendon Press.

\end